\documentclass{aa}
\usepackage[dvips]{graphicx}
\def\simless{\mathbin{\lower 3pt\hbox
{$\rlap{\raise 5pt\hbox{$\char'074$}}\mathchar"7218$}}}   
\def\simmore{\mathbin{\lower 3pt\hbox
{$\rlap{\raise 5pt\hbox{$\char'076$}}\mathchar"7218$}}}   
\newcommand{\be}{\begin{equation}}
\newcommand{\ee}{\end{equation}}

\newcommand{\rd}{{\rm d}}
\begin{document}

\title{Spectra of Poynting-flux powered GRB outflows}
\titlerunning{Spectra of Poynting-flux powered GRB outflows}
\authorrunning{Giannios \& Spruit}
\author{Dimitrios Giannios \inst{1,2,3}
\and Henk C. Spruit \inst{1}
}

\institute{Max Planck Institute for Astrophysics, Box 1317, D-85741 Garching, 
Germany
\and University of Crete, Physics Department, P.O. Box 2208, 710 03, 
Heraklion, Crete, Greece
\and Foundation for Research and Technology-Hellas, 711 10, Heraklion, 
Crete, Greece
}

\offprints{dgiannios@physics.uoc.gr}
\date{Received / Accepted}

\abstract{

We investigate the production of the gamma-ray spectrum of a Poynting-flux dominated 
GRB outflow. The very high magnetic field strengths (super-equipartition)
in such a flow lead to very efficient synchrotron emission. In contrast 
with internal shocks, dissipation of magnetic energy by reconnection is gradual
and does not produce the spectrum of cooling electrons associated with shock
acceleration. We show that a spectrum with a break in the BATSE energy range 
is produced, instead, if the magnetic dissipation heats a small ($\sim 10^{-4}$)
population of electrons.
}

\maketitle

\section{Introduction} \label{intro}

Magnetic fields are an attractive ingredient in central engine models of
$\gamma$-ray bursts because of their ability to transfer large amounts of energy
across (near) vacuum and depositing it into a small amount of matter. The energy
flux in such magnetically powered, or `Poynting-flux dominated' outflows (refs
Thompson 1994; Lyutikov \& Blandford 2003; Drenkhahn 2002) is initially in
electromagnetic form. To produce a GRB, this energy flux has to be transferred to
the matter in such a way as to produce both  bulk flow speed and the right
amount of nonthermal radiation.

Poynting-flux dominated outflows come in 2 basic types: the axisymmetric or `DC'
models and the nonaxisymmetric or `AC' ones (for discussions of these models
see Lyutikov \& Blandford 2003; Drenkhahn 2002 [Paper I]; Spruit \& Drenkhahn 
2002 [Paper II]). An axisymmetric, rotating magnetic field can accelerate flows by the
magnetocentrifugal mechanism. Instabilities in the outflow can dissipate some of
the magnetic energy to produce heating, acceleration of fast particles and
ultimately nonthermal radiation. In the AC model on the other hand, the rotating
magnetic field of the central engine is nonaxisymmetric, producing an outflow in
which the field lines change directions rapidly, at the rotation frequency of the
central engine or higher. Dissipation of magnetic energy by reconnection in this
flow is much faster than in the DC model, and the acceleration of the outflow
very efficient, though by a somewhat different mechanism than in the
magnetocentrifugal process (Papers I; II, for a
summary see Spruit \& Drenkhahn 2003).

In both these models, the observed nonthermal radiation is produced by
dissipation of magnetic energy by reconnection. Details of the
reconnection process matter for the radiation produced, but are rather poorly
understood at present. Significant parts of the relevant plasma physics are not
known, especially under the relativistic conditions of a GRB. This hampers for the
moment the development of models for the prompt radiation to be expected from
a magnetically powered GRB outflow. Nevertheless, a few reasonably established
facts already help in limiting the possibilities.

In this paper we develop the consequences of assuming magnetic reconnection as
the source of gamma-radiation from a GRB. Like the internal shock model, the
model has a free parameter representing the fraction of the electrons that takes
up the dissipated energy and produces the synchrotron radiation. It has one
degree of freedom less, however, since the magnetic field strength in the flow
follows within narrow limits from the energetics, once the choice for an AC or DC
model is made. In the developments below, an AC model is assumed for
quantitative results.

An important diagnostic of the prompt emission is the break in the Gamma-ray
spectrum. This break is usually assumed to represent the synchrotron spectrum from
a distribution of electrons with a lower energy cutoff, as is produced by the
synchrotron cooling of quasi-instantaneously accelerated electrons. This may be
appropriate for an internal shock model. The idea has also been carried over to
magnetic prompt emission models by Sikora et al. (2003). It turns out, however,  that
magnetic reconnection is not nearly instantaneous enough to produce such a
distribution. In magnetic reconnection, the electron distribution will reflect, instead,
the local instantaneous balance between synchrotron cooling and heating by the
reconnection-related electric fields.

\section{Magnetically powered outflows}

The outflow carries free energy in the form of wound-up magnetic field lines, 
which can be extracted through reconnection processes.
For reconnection processes to take place,
differently oriented field lines must come close to each
other. The length scale on which the orientation of
magnetic field lines changes controls the speed of field
dissipation; the smaller the length scale the faster the field decay.  In DC
models, the field lines are in the same direction, and dissipation has to wait
until global instabilities have caused changes in direction between nearby field
lines. As a result the dissipation in such models tends to happen at a rather large 
distance, of order $10^{17}$ cm (Lyutikov \& Blandford 2003).

We consider here the outflow from a nonaxisymmetric
rotator. This configuration gives rise to a `striped' wind (Coroniti
1990, see also Spruit et al. 2001), where the magnetic field varies
with a wavelength of $\lambda \sim 2\pi c/ \Omega$ (in the source's frame).   
A detailed description of the model has been given previously (Papers I 
and II). In this section we summarize the main ingredients of the
model and its assumptions.

\subsection{Reconnection in a relativistic magnetic field}
Both the dynamics of the flow and the radiative energy release are governed by
the rate of magnetic reconnection in the flow. The magnetic energy flux is of the
same order as the kinetic energy flux $\Gamma \dot m c^2$ where $\dot
m=\dot M/4\pi$ is the (baryonic) mass flux per steradian and $\Gamma $ the
bulk Lorentz factor. In a frame comoving with the flow, the magnetic energy
density is then of the same order as the rest mass energy density, i.e. the field is
relativistic (Alfv\'en speed of the order of the speed of light) and highly {\em
super-equipartition}.  

 The outflow speed of the plasma from a reconnection region is
of the order of the Alfv\'en speed, rather independent of the
particular reconnection model. Under relativistic conditions
(energy density in the magnetic field larger than the rest
mass energy density), Blackman \& Field (1994a) (see also Blackman 
\& Field 1994b) find that the inflow speed into the reconnection layer is 
increased by the relativistic kinematics, and is also subrelativistic.
This has also been found by Lyutikov \& Uzdensky (2003). Thus
we may assume that the velocity $v_{\rm rec}$ at which the magnetic energy
density $B^2/8\pi$ flows across a reconnection boundary is of the order of the
speed of light
\be
v_{rec}=\epsilon v_{\rm A},
\ee
where $v_{\rm A}\approx c$ and $\epsilon$ is of order unity. 
We keep this parameter in the calculations, but consider it a relatively well determined
quantity, conservatively set at $\epsilon=0.1$ for quantitative estimates.

\subsection{The dynamics of the flow}
\label{dyn}

Near the source the
flow is accelerated magnetocentrifugally up to a distance around the
Alfv\'en radius. Then it becomes asymptotically radial. The poloidal and
azimuthal components are similar in magnitude; however, from that
point on their magnitude scales as $B_{\phi}/B_r \sim r$, resulting in a
negligible radial component at larger distances. In Poynting-flux
dominated outflows the Alfv\'en radius is almost equal to the light
radius $c/\Omega$. So for distances $r>c/\Omega$ the flow can be
assumed to be purely radial with a dominant azimuthal component of the
magnetic field.

From this point on we focus on the dynamics of the flow, as well as on
its radiative properties at distances $r>r_0 \sim$ a few $c/\Omega$.   
An important parameter of the model is the ratio ${\sigma}_0$ of the 
Poynting flux to the kinetic flux at the initial radius
\be
{\sigma}_0=\frac {L_{\rm B,0}}{L_{\rm kin,0}}=\frac {\beta_0 (B_0
r_0)^2}{4 \pi {\Gamma}_0 \dot M c},
\label{sigma}
\ee
where $\beta=v/c$, $B_0$ and ${\Gamma}_0$ are the magnetic field
strength  and the bulk Lorentz factor at the radius $r_0$ and $\dot M$
is the mass flux per sterad. The flow starts at $r_0$ with the Alfv\'en
speed (${\Gamma}_0=\sqrt {{\sigma}_0+1}$).
In the GRB case the flow must start highly Poynting-flux dominated
with $\sigma_0 \simmore 100$ so that $\Gamma \gg 1$ at all distances. 
So we can safely set ${\sigma}_0+1\simeq {\sigma}_0$ and $\beta
\simeq 1$.

There is also a simple relation that links the total luminosity of the
flow to the mass flux
\be
L={\sigma}_0 {\Gamma}_0 \dot M c^2.
\label{link}
\ee
A characteristic value of the asymptotic Lorentz factor of the flow is
\be \Gamma_\infty\approx\sigma_0 \Gamma_0 . \label{gammi}\ee
Its actual value will differ, depending on which fraction of the Poynting flux is
converted to kinetic energy. In the AC model, this is about 50\% if most of the
dissipation takes place outside the photosphere, and larger (up to 100\%) when
dissipation is near or inside the photosphere.

We will apply the results to cases where magnetic dissipation is well outside the
photosphere so that the asymptotic kinetic luminosity  is similar to the initial
Poynting flux.  In the region $r<r_{\rm s}$  where the Poynting flux has not yet
dissipated significantly, this implies the approximate relation
\be {B^2\over 8\pi}c\approx \Gamma_\infty\rho c^3,\ee
in the frame of the central engine, or
\be  
{B^{\prime 2}\over 8\pi}\approx {\Gamma_\infty \over\Gamma} 
\rho^\prime c^2,\label{bpri}\qquad (r\simless r_{\rm s})
\ee
in a frame comoving with the bulk flow (indicated by primes). Thus, as mentioned
above, the magnetic field in the outflow is relativistic, up to the saturation 
radius $r_{\rm s}$ where most of the dissipation takes place.
 
With magnetic dissipation scaling with the Alfv\'en speed as discussed
above, the acceleration of the flow can be computed in detail,
including the gas and radiation pressures where relevant. The results
of detailed numerical calculations where reported in Paper II. The 
acceleration of the flow stops when most of the free magnetic energy has 
been dissipated; this happens at a distance $r_{s}$. In the intermediate range 
of radii $r_0\ll r\ll r_{s}$ 
the equations simplify and an analytic solution is possible (Paper I). The
analytic solution has been found to be fairly close to the numerical
results and will be used in this work so as to make predictions on the
radiative properties of the flow.

In the intermediate range of radii the bulk Lorentz factor of the flow
scales as $\Gamma \propto r^{1/3}$, while beyond the saturation
radius it can be approximated as constant
\be
\Gamma=\Gamma_\infty(r/r_{\rm s})^{1/3}
\quad \rm {for}\quad r<r_{s},
\label{bulk}
\ee
\be
\Gamma = {\Gamma}_{\infty}={\sigma}_0^{3/2} \quad \rm {for}
\quad r>r_{s},\label{gam}
\ee  
where 
\be r_{\rm s}= \pi c\Gamma_\infty^2/(3\Omega\epsilon), \label{satur} \ee
or, in terms of the light cylinder radius $r_{\rm l}$,
\be r_{\rm s}/r_{\rm l}={\pi\over 2\epsilon}\Gamma_\infty^2.\ee
For canonical GRB parameters ($\Gamma=300$, $\epsilon=0.1$, $r_{\rm l}=10^7$
cm, it is at $r_{\rm s}\sim 10^{13}$ cm, a factor 10-100 outside the Thomson
photosphere. 

We can see how much magnetic energy is dissipated in the flow (in the central
engine's frame) as a function of radius following the Poynting flux in the flow.
The latter is given by the expression (Paper I)
\be
L_{\rm B}=\beta c
\frac{(rB)^2}{4\pi}=L(1-\frac{\Gamma}{{\sigma}_0^{3/2}}),
\label{pf}
\ee
where $L$, $L_{\rm B}$ are the the total flux and the Poynting flux of the flow
per steradian.
From Eq. (\ref{pf}) and Eq. (\ref{bulk}) one can see that the Poynting
flux in the flow drops as a function of distance. This reduction
corresponds to the magnetic energy that is gradually dissipated in the
flow. The energy rate $d\dot E_{\rm rel}$ that is released in a shell ($r,r+\rd r$) 
of the flow will be
\be
\rd\dot E_{\rm rel}=-\frac{\rd L_{\rm B}}{\rd r}\cdot \rd r=
\frac{0.049L}{r_{11}^{2/3}}
\frac{(\epsilon\Omega)_3^{1/3}}{{\sigma}_{0,2}}\rd r_{11},
\label{de}
\ee
where $r=10^{11}r_{11}$ cm.
So in the frame of the central engine $ \rd\dot E_{\rm rel}\propto
\rd r/r^{2/3}$. After integration in a range of radii we find $\dot E\propto
r^{1/3}$. So, most of the energy is actually dissipated in the outer
regions of the flow, close to the saturation radius.
 
Other physical quantities of the flow that will be important are the proper 
density $\rho$, the magnetic field strength $B$, and the location
of the Thomson photosphere
$r_{\rm ph}$.
Combining the continuity equation $\dot M=r^2\Gamma \rho c$ with 
Eq. (\ref{link}) and Eq. (\ref{bulk}) we have for the proper density
\be
\rho = \frac {\dot M}{r^2 \Gamma c}=\frac{L}{{\sigma}_{0,2}^{3/2} r^2
\Gamma c^3}= \frac {2.5\cdot 10^{-7}}{r_{11}^{7/3}}
\frac{L_{52}}{{\sigma}_{0,2}^2 (\epsilon \Omega)^{1/3}}  \quad \rm
{g/cm^3},
\label{density}
\ee
where $\dot M$ is the mass flux per steradian and $L= 10^{52} L_{52}$
 erg/sec/sterad is a typical value for the
observed GRB luminosities taking into account the strong evidence for
highly collimated outflows (Frail et al. 2001). 

Using Eq. (\ref{pf}) and Eq. (\ref{bulk}) (for $\Gamma\ll
{\Gamma}_{\infty}$) the comoving magnetic field strength will be
\be
B^\prime =B/\Gamma = \frac{25\cdot 10^7}{r_{11}^{4/3}}\frac
{L_{52}^{1/2}}{(\epsilon \Omega)_3^{1/3} {\sigma}_{0,2}^{1/2}} \quad
\rm {Gauss}.
\label{Bfield}
\ee
The radius of the photosphere of the flow depends on the baryon loading and 
the energy flux. For large bulk Lorentz factors it is given by integrating 
the expression (cf. Abramowicz et al. 1991)
$\rd\tau = \Gamma (1-\beta) k_{es} \rho \rd r$ from $r$ to $\infty$ to find
\be 
\tau = \frac{20}{r_{11}^{5/3}}
\frac{L_{52}}{{\sigma}_{0,2}^{5/2}(\epsilon \Omega)_3^{2/3}}.
\label{tau} 
\ee
Setting $\tau(r_{\rm ph})=1$ and solving for the location of the photosphere, we
find its position as a function  of the parameters of the flow
\be
r_{ph,11} =6 \frac{L_{52}^{3/5}} {{\sigma}_{0,2}^{3/2}(\epsilon
\Omega)_3^{2/5}}. 
\label{r_ph}
\ee

\section{The reconnection layer}
\subsection{Time scales}

Let the reconnection layers have  a lateral size $L$ (given by the wavelength 
of the field reversals) and an as yet unknown thickness $\delta$. Matter 
and magnetic field lines are advected to the reconnection
region with the reconnection velocity $v_{\rm rec}$ through the upper and lower
surfaces of the layer. The material is pulled out of the reconnection region
through the sides of the layer with speeds comparable to the Alfv\'en speed 
by magnetic tension forces. While $\delta$ depends strongly on the poorly
known reconnection physics, the outflow speed is approximately known, since
the Alfv\'en speed is of the order of the speed of light, as noted above. The 
width $L$ of the reconnection layer is  given by the intrinsic
length scale in the flow, namely the wavelength of field reversals. 
Hence, in a frame comoving with the bulk flow $L\approx \Gamma \pi 
c/\Omega$. The residence time of matter in the reconnection layer is therefore 
of the order
\be \tau_{\rm r}= L/v_{\rm A}\approx \Gamma \pi/\Omega=\pi r_{\rm
l}/c\ee
where $r_{\rm l}$ is the light cylinder radius of the central engine. 

This time scale can be compared with the cooling time of the electrons in the
reconnection layer. On account of the very high field strengths, the dominant
radiation process is most likely synchrotron emission. Because the details of the
particle acceleration processes in the reconnection layer are unknown, we assume
that the energy dissipated goes mostly into a fraction  $\xi\ll 1$ of the electrons,
which as a result attain a typical Lorentz factor $\gamma$.  The parameter $\xi$
plays the same role as its counterpart in the internal shock model. Note, however,
that there is no additional parameter for the strength of the magnetic field, since this
is fixed by the Poynting flux model.

If the characteristic Lorentz factor of the radiating electrons is such that the
radiation seen in the observer's frame has energy $h\nu_{\rm obs}/m_{\rm e}
c^2= x_{\rm obs}\approx 1$, the cooling time of the electrons (in a comoving frame)
can be written as
\be 
\tau_{\rm s}=6\pi\sigma_{\rm T}^{-1} \Gamma^{1/2}x_{\rm obs}^{-1/2} 
(e\hbar/c)^{1/2} B^{\prime -3/2}. 
\ee
The electron cooling time is very short due to the high magnetic field strengths
that make the electrons radiate efficiently. The electrons thus stay cold, and if
most of the magnetic energy is dissipated to the electrons, the flow as a whole 
stays cold.

The reconnection time scale $\tau_{\rm r}$ is of the order of the light travel time
over the length scale of the magnetic field in the comoving frame
\be \tau_{\rm r}\approx \pi r_{\rm l}\Gamma/c.\ee
In terms of the energy flux in the burst, and making use of Eq. (\ref{gam}), the ratio
of the two time scales becomes
\be 
\tau_{\rm s}/\tau_{\rm r}\approx 10^{-7}(r_{\rm l}/10^7)^{1/2}x_{\rm 
obs}^{-1/2}(r/r_{\rm s})^{11/6} L_{52}^{-3/4} (\epsilon/0.1)^{-3/2}.
\ee

The synchrotron cooling time is thus very short compared with the residence time
of the radiating electrons in the acceleration region. This has the consequence
that the usual approximation of sudden acceleration followed by cooling of the
electron population made in the internal shock model is {\it not} a good model
for the synchrotron emission from magnetic reconnection. Rather, the electron
energy distribution must be close to an {\it equilibrium} between synchrotron
cooling and heating through reconnection.

\subsection{Energetics of the reconnection region}

 In this section we estimate the rate at which magnetic energy is released 
in a reconnection layer. Matter and magnetic field lines are advected through 
the upper and lower surfaces of the layer with the reconnection velocity $v_{rec}$. 
The flux of the incoming magnetic energy in the reconnection layer is
\be 
F_{\rm B}=2\frac{B'^2}{8\pi}v_{\rm rec},
\ee
where the factor 2 accounts for the two sides of the layer. This magnetic flux
is dissipated over the thickness $\delta$ of the layer resulting in a rate of
magnetic energy density released
\be
\dot E_{\rm dis}=\frac{B'^2}{4\pi}\frac{v_{\rm rec}}{\delta}.
\label{dissip}
\ee 
The material leaves the reconnection region
through the side of the layer with speeds comparable to the Alfv\'en
speed. If the proper density of the flow is ${\rho}_{\rm in}$ and
${\rho}_{\rm out}$ inside and outside the layer respectively, then 
mass conservation gives
\be
L {\rho}_{\rm out} v_{\rm rec}{\Gamma}_{\rm rec}= \delta {\rho}_{\rm
in} v_{A,co} {\Gamma}_{A,co},
\label{layer}
\ee
where $ v_{A,co}$ is the comoving Alfv\'en speed, with its corresponding
Lorentz factor
\be
 {\Gamma}_{\rm A,\rm co}=\sqrt{1+\sigma}, \quad \sigma=\frac {B_{\rm
co}^2}{4\pi \rho c^2}=\frac {6.8}{r_{11}^{1/3}}\frac
{{\sigma}_{0,2}}{(\epsilon \Omega)_3^{1/3}}.
\label{v_A}
\ee      
For typical distances of interest $r_{11}\simmore 0.1$, ${\Gamma}_{A,co}\sim$ a
few. Finally, $\Gamma_{\rm rec}$ is the Lorentz factor that corresponds to the 
reconnection velocity. 
Since $v_{\rm rec}=\epsilon v_{\rm A,co}$, and the reconnection speed parameter 
$\epsilon$ is not close to unity, we have $\Gamma_{\rm rec}\simeq 1$.

Mass conservation (Eq. (\ref{layer})) provides a constraint on $\delta$, 
but something has to be known about the density ratio $\rho_{\rm out}/
\rho_{\rm in}$ for this to be useful. This in turn depends on the efficiency 
with which the dissipated energy is radiated away. If the energy stays in 
the particles, then the
particle pressure can compensate the decline of the magnetic
energy pressure in the dissipation layer and the flow is essentially
incompressible. If, however, the particles radiate very efficiently,
they stay relatively cool, resulting in a highly compressible flow
(i.e. ${\rho}_{\rm in}\gg {\rho}_{\rm out}$). 
 Thus, the thickness of the reconnection layer remains somewhat
uncertain, but it turns out that this has no effect on the spectral estimates
made in the next section, since the value of $\rho_{\rm in}$ drops out.

For parameters of the model relevant to GRBs, the protons are inefficient radiators 
owing to their high mass in contrast to electrons. 
It is often argued that in reconnection most of the magnetic energy is dissipated
into the electrons rather than the ions, on account of their greater mobility in the
electric fields generated in the reconnection region. It is not known to what
extent this still holds for the relativistic reconnection environment of a GRB. For
completeness we keep the fraction dissipated into the electrons, $f_e$ as a free
parameter. If  $f_e\ll 1$, most of the energy is transfered to the protons and,
unless there is a very efficient mechanism of energy coupling between electrons
and protons, most of the energy will serve to accelerate the bulk flow through
adiabatic cooling of the protons. In this case the radiative efficiency of the flow
becomes too low to explain a GRB.  As in the internal shock model, we assume 
that this is not the case, i.e. $f_{\rm e}$ is taken of order unity.

Another parameter of the model is the fraction $\xi$ of the electrons
that will be accelerated in the reconnection region. The two parameters
 $f_e$ and $\xi$ correspond to the same degrees of freedom as in the 
internal shock model, where a large fraction of the energy that is dissipated in the
shocks accelerates the electrons, and another fraction is assumed to build
random magnetic fields. 
  
 We can estimate the characteristic Lorentz factor to which
the electrons are accelerated in the reconnection layers at different
distances from the central engine. The electrons are heated at a rate
\be
\dot E_{\rm heat}=f_e \dot E_{\rm dis}.
\label{heating}
\ee
An electron moving with a Lorentz factor ${\gamma}_{\rm ch}$ in a 
magnetic field of strength $B^\prime$ will radiate at a rate ${\sigma}_T
v^2{\gamma}_{\rm ch}^2 B^{\prime 2}/(6\pi c)$. If ${\gamma}_{\rm ch}$ is 
the characteristic Lorentz factor of the electron distribution then the cooling 
rate per unit volume due to synchrotron loss in the optical thin regime
will be 
\be
\dot E_{\rm cool}={\xi\rho_{\rm in}\over m_p}{\sigma_{\rm T}
v^2\gamma_{\rm ch}^2 B^{\prime 2}\over 6\pi c}.
\label{cooling}
\ee 

Inside the reconnection layer electrons are accelerated until $\dot
E_{\rm heat}=\dot E_{\rm cool}$. Using Eq. (\ref{heating}),
Eq. (\ref{cooling}) and  Eq. (\ref{dissip}) we find
\be 
{\beta}_{\rm ch} {\gamma}_{\rm ch} = 7\cdot r_{11} \frac 
{{\sigma}_{0,2}^{3/4} {\Omega}_4^{1/2}}{{\xi}^{1/2}_{-4}L_{52}^{1/2}},
\label{gammae}
\ee
where we have also set $f_e {\Gamma}_{A,co}=1$ (i.e. $f_e\simeq
0.5$) and  ${\rho}_{\rm out}=\rho$ (Eq. \ref{density}).
We have also normalized $\xi=10^{-4}\xi_{-4}$.

Equation (\ref{gammae}) shows that the characteristic energy of the electrons
actually turns out  to be independent of the uncertain thickness $\delta$
of the reconnection region (related to the effective compressibility of the
reconnection flow). This can be understood by the following
consideration: high compressibility means small thickness $\delta$ of the
reconnection layer and more energy dissipation per unit volume in the
reconnection sites. On the other hand, this energy has to be shared
among more particles (since the density is higher), resulting in the
same heating rate per particle.

\section{The spectrum}
In this section we consider the radiation from reconnection regions in the
outflow, under the assumptions that i) the dissipated energy goes into a fraction
$\xi$ of the electrons, and ii) the electrons thus heated have a narrow or
quasi-thermal distribution. The spectrum of the burst is then the synchrotron
spectrum of this distribution, folded over the distance from the source and
Doppler shifted to the observer using the radial dependences given in Sect.
\ref{dyn}.

\subsection{ The peak frequency of the spectrum}

Equation (\ref{gammae}) is simplified in the ultrarelativistic regime, where 
${\gamma}_{\rm ch}\propto r$. The characteristic photon energy at which
the electrons emit (in the comoving frame) scales as ${\nu}^\prime\propto 
{\gamma}_{\rm ch}^2 B^\prime \propto r^{2/3}$ and is Doppler boosted 
in the central engine's frame to an energy ${\nu}_{\rm ch} \propto \Gamma
{\nu}^\prime \propto r$. So the most energetic photons are expected
to be emitted in the outer parts of the flow close to the saturation
radius. Since this is also where most of the magnetic energy is dissipated, these
photons will give rise to the peak of the $\nu \cdot f_{\nu}$ spectrum.    

More quantitatively, in the central engine's frame the characteristic
frequency emitted as a function of radius is ${\nu}_{ch}= 3/2 \Gamma
{\gamma}_{\rm ch}^2{\nu}_{\rm L}$, where ${\nu}_{\rm L}=eB^\prime/
(2\pi mc)$ is the Larmor frequency. Using Eq. (\ref{Bfield}), Eq. (\ref{bulk}),
and Eq. (\ref{gammae}) we get in the relativistic limit
\be
X_{\rm ch}={h\nu_{\rm ch}\over mc^2} \approx 6\cdot 10^{-3}
r_{11} {\Omega_4 (\Gamma_\infty/300) \over \xi_{-4}L_{52}^{1/2}}.
\label{X_obs}
\ee
$X_{\rm ch}$ becomes maximum at the saturation radius
$r_{s}$. Combining  Eq. (\ref{X_obs}) and Eq. (\ref{satur}):
\be
X_{\rm ch,max}= 
0.3 \frac{(\Gamma_\infty/300)^3}{\xi_{-4}{\epsilon}_{-1}L_{52}^{1/2}}.
\label{xe}
\ee

The peak in the $\nu \cdot f_{\nu}$ spectrum, which
corresponds to the break of the Band spectrum (Band et al. 1993), is in the
range $X_{\rm br} \sim (0.1-1)$. In the central engine's frame
(assuming a typical redshift $\sim 1$) the same peak is in the
range $\sim (0.2-2)$. In the case that all the electrons are accelerated
in the reconnection region ($\xi=1$), if the peak is to be in the observed
range, we must either have  $\epsilon\ll 0.1$, or $\Gamma_\infty \simmore
1000$.

If, however, only a small fraction ($\xi\sim 10^{-4}$) of the
electrons is accelerated, values $\epsilon\simeq 0.1$ and $\Gamma_\infty\simeq
300$ are compatible with the characteristic of the prompt radiation. When specific
values of the parameters have to be chosen for arithmetic  examples, these values 
will be used in what follows.

\subsection{Spectral energy distribution}

For a prediction of the spectrum the actual electron distribution is needed,
 not only a characteristic Lorentz factor. The electron distribution depends on
the acceleration mechanisms that take place in the reconnection layers
and on the cooling of the electrons. A thermal distribution
is expected when particles can exchange energy efficiently. For
the conditions expected in the reconnection layer, the heating and
synchrotron cooling time scales can be shown to be orders of
magnitude shorter than the energy exchange time scale due to Coulomb
collisions. So Coulomb collisions are inefficient at thermalizing the
electrons in the reconnection sites.

Another thermalization mechanism is synchrotron absorption. When the
bulk of the radiating energy of the particles is reabsorbed, this provides an
efficient channel of energy exchange between the particles, and a
thermal distribution is expected. The role of the synchrotron
absorption will be explored in the next section, where it will be
shown that at distances comparable to or larger than the Thomson
photosphere, electrons emit mostly in an environment optically thin to
synchrotron radiation. 

Under these conditions, at radii larger than the photospheric radius
non-thermal electron distributions may appear. As an illustrative
example we assume that the accelerated electrons have a power law
distribution $N(\gamma)\propto {\gamma}^{-p}$, with a lower cutoff 
at the characteristic Lorentz factor as defined in
Eq. (\ref{xe}) and $p> 2$. The spectrum is obtained by integration over this
distribution and over all contributing radii (those outside the photosphere),
taking into account the increase of the Doppler shift with distance (Eq.
(\ref{bulk})).  An example of this spectrum is shown in Fig. 1.                   
 
The figure shows that the peak in the $\nu\cdot f_{\nu}$ spectrum 
is indeed the characteristic emitted frequency at the saturation radius
(Eq. (\ref{xe})). The slope of the spectrum above the peak $\beta$ is 
determined by the power law of the electron distribution through the relation 
$\beta=-(p-1)/2$. Below the peak the spectrum shows some curvature (it
cannot be exactly fitted with a power law) and is a result of the
superposition of the contributions from all the radii below the saturation
radius.

\begin{figure}
\resizebox{\hsize}{!}{\includegraphics[]{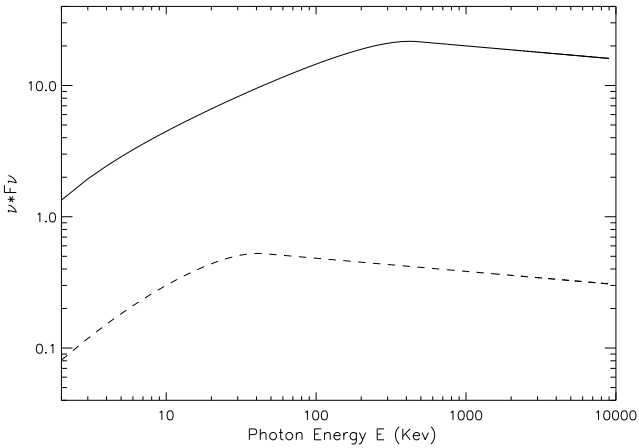}}
\caption[]{Illustrative example of the spectrum (in the source's
frame) emitted by a power law ${\gamma}^{-p}$ distribution of electrons in the
reconnection centers. In this example $p=3.2$ resulting in the high energy slope of
the spectrum. The other parameters are $\epsilon=0.1$, $\Omega= 10^4$
rad/sec, ${\sigma}_0=100$, $L=3.2\cdot10^{51}$ erg/sec/steradian and $\xi=
2\cdot10^{-3}$. The dashed line corresponds to $L=10^{50}$
erg/sec/steradian	 and ${\sigma}_0=40$, parameters typical for X-ray flashes.}  
\label{fig1}
\end{figure}

\section{X-ray flashes}

The origin of X-ray flashes and their connection to GRBs is not clear
yet. However, recent analysis (Barraud et al. 2003) has indicated that
X-ray flashes actually belong to the same family as GRBs, showing
similar spectral characteristics. X-ray flashes appear to be dimmer
than GRBs and their peak in the $\nu\cdot f_{\nu}$ spectrum is at a photon 
energy $E_{\rm m}$ of less than 50
keV (this is actually one way to define X-ray flashes). Barraud et al
argue that X-ray flashes may occur at redshifts comparable to those of GRBs
since their typical duration does not indicate stronger cosmological
time dilation effects (their duration is similar to this of GRBs).

Furthermore, Amati et al. (2002) found indications that $E_{\rm m}$ 
correlates with the square root of the equivalent isotropic
luminosity released in the GRBs  ($L\sim {X^{1/2}_{\rm ch,max}}$ in our
notation). If we assume that this correlation may be extended to the case of
X-ray flashes, then a typical X-ray flash, with $E_{\rm m}$ ten times
lower than that of a GRB, would have two orders of magnitude less isotropic 
luminosity. 
      
So it seems that a more appropriate normalization value of the
luminosity per steradian for the typical X-ray flash is
$L=10^{49.5}=L_{49.5}$ erg/sterad/sec. In terms of this model a peak in
the spectrum of $\sim20$ keV (which corresponds to $\sim 40$ keV in
the central engine's frame for $z\sim 1$) and using Eq. (\ref{xe}), we
find that $\sigma_{0} \simeq 40$. Here we have kept $\epsilon=0.1$,
$\Omega =10^4$ rad/sec and $\xi\sim 10^{-3}$. 

The previous argumentation indicates that a wind which is less Poynting-flux 
dominated (i.e. with higher baryon
loading) can give rise to an X-ray flash. For these parameters the
saturation radius (Eq. (\ref{satur})) is $r_{\rm s}\simeq 2\cdot
10^{12}$ cm, while the photosphere is located at $r_{\rm ph}\sim 
 10^{11}$ cm. For $r_{\rm ph}<r<r_{\rm s}$ the radiation emitted
has a similar spectrum to that of GRBs (see Fig. 1), but with a peak shifted
in the X-ray region (at $\sim 20$ keV). An important fraction of the
total energy is, however, dissipated below the photosphere, where
scattering plays an important role (see Sect. \ref{photo}).         

\section{The role of synchrotron absorption in the reconnection layers}

The shape of the energy spectrum emitted from the reconnection sites also 
depends on the synchrotron absorption. The optical depth due to synchrotron 
absorption is frequency-dependent and at low enough photon energies the
medium becomes optically thick and the spectrum hardens. The photon frequency
at which the medium makes a transition from the synchrotron thick region to the
synchrotron  thin region is called the turn-over frequency $\nu_{t}$. In this
section we calculate the turn over frequency as a function in the flow and
examine
its implications for the emitted spectrum. 

An electron of Lorentz factor $\gamma$ in a
magnetic field of strength $B^\prime$ has a frequency-dependent cross section 
for synchrotron absorption given by
(Ghisellini \& Svensson 1992)
\be 
{\sigma}_s(\nu,\gamma)=\frac{3\pi}{10}\frac{{\sigma}_{T}}{{\alpha}_{f}}
\frac{B_{\rm cr} }{B^\prime}\frac{x}{\gamma^5}\Big(K_{4/3}^2(x/2)-
K_{1/3}^2(x/2)\Big),
\label{sigmas}
\ee
where $B_{\rm cr}\simeq 4.4 \cdot 10^{13}$ Gauss, ${\alpha}_f$ is the
fine structure constant, $x=2\nu/(3\gamma^2{\nu}_L)$,
${\nu}_L=eB^\prime/(2\pi mc)$ and $K_{4/3}$, $K_{1/3}$ are modified Bessel
functions.
This expression is valid for $\gamma\gg 1$ and for an isotropic
distribution of electrons. 
Integrating this over a power law distribution of electrons
$N(\gamma)\propto {\gamma}^{-p}$, we find the absorption coefficient
\be
{\alpha}_{\nu}=\int_{{\gamma}_{\rm ch}}^{{\gamma}_{\rm max}}
{\sigma}_s(\nu,\gamma) N(\gamma)d\gamma
\label{a_v}
\ee       
where ${\gamma}_{\rm ch}$ is given by Eq. (\ref{gammae}) and ${\gamma}_{\rm
max}\gg {\gamma}_{\rm ch}$. 

The width of the layers (in the comoving frame) is $\delta$ and the turn
oner frequency ${\nu}_{\rm t}$ for
synchrotron absorption can be estimated solving the equation
\be
{\alpha}_{{\nu}_{\rm t}}\cdot \delta=1.
\label{v_t}
\ee 
The thickness of the reconnection layer $\delta$ is related to its typical length $L$
through Eq. (\ref{layer}). This estimate is rather rough; fortunately it turns out
that the turn-over frequency depends very weakly on the actual dimensions of the 
layer.

For the estimate we use as typical parameter values $p\simeq 3,\quad
\Gamma_\infty =1000,\quad {\Omega}=10^4$ rad/sec, $L=3.2\cdot10^{51}$
erg/sec/steradian and $\xi=2\cdot10^{-3}$.  
Most of the radiation is emitted where most of the energy is dissipated, i.e. 
near the saturation radius $R_{\rm s}$ (Eq. (\ref{satur})). At this distance
the turn-over frequency 
lies in the soft X-rays ($\nu_{t}\sim 300$ eV). At smaller radii, the turn-over
frequency increases and in the location of the Thomson photosphere it is 
around $\sim 10$ keV. Thus synchrotron self-absorption can play
a significant role in the shape of the electromagnetic spectrum in the soft
X-ray region, making the spectra harder than what is plotted at the low 
($E<10$ keV) energy end of Fig. 1.    
	 
\section{Below the photosphere}
\label{photo}

Until now, we have explored the radiative properties of the flow in the 
Thomson-thin region (i.e. $r> r_{\rm ph}$). In this section we briefly discuss the 
energetics of the quasi-thermal component expected from photons generated
below the photosphere. We will call this the photospheric component.  

At small radii, or equivalently at large optical depths, radiation and matter are
expected to be in thermodynamic equilibrium, sharing the same (comoving)
temperature $T$. The thermal energy density, which is dominated by radiation, is
fed by the dissipated magnetic energy and suffers from adiabatic losses at the
same time. The result is that only a fraction of the energy that was initially
injected in thermal form will appear as black-body radiation when matter and
radiation decouple.

The power dissipated as a function of  distance is given by Eq. (\ref{de}). As
shown in paper II, half of this energy accelerates the flow, and the other half is
injected into the flow as thermal energy. The scaling of
this energy rate with radius is $d\dot E\propto \rd r/r^{2/3}$ (Eq. (\ref{de})). 
Suppose now that energy is released at some rate $L_{\rm inj}$ at a radius
$r$ and we want to know how much of it will appear as thermal radiation
when matter and radiation decouple at the photosphere. For adiabatic
expansion of a radiation-dominated flow, the pressure scales as $\pi\propto
\rho^{4/3}$ and $\pi\propto T^4$. Combining these we have 
$T\propto \rho^{1/3}\propto r^{-7/9}$, where in the last step 
Eq. (\ref{density}) has been used. 

With the scalings $\Gamma\propto r^{1/3}$ (Eq. (\ref{bulk})) and $T\propto 
r^{-7/9}$, we have for the luminosity: $L(r) \propto
r^2{\Gamma}^2T^4\propto r^{-4/9}$. So, only a fraction $(r/r_{\rm
ph})^{4/9}$ of the initial thermal energy will still be in thermal
form at $r_{\rm ph}$. So a shell at distance $r, r+\rd r $  contributes to
the luminosity emitted at $r_{\rm ph}$
\be
\rd L\propto \frac{\rd r}{r^{2/3}}(\frac{r}{r_{\rm ph}})^{4/9}.
\ee
Integrating the last expression from 0 to $r_{\rm ph}$ we find for the 
photospheric luminosity
\be
L_{\rm ph}=3\cdot 10^{50} L_{52}r_{\rm ph,11}^{1/3}\frac{(\epsilon
\Omega)_3^{1/3}}{{\sigma}_{0,2}} \qquad \rm erg/sec/sterad.
\label{lbb}
\ee 
Where $r_{\rm ph}$ is given by Eq. (\ref{r_ph}). 

For the typical GRB parameters, one can check that the photospheric component
is about one order of magnitude weaker than the total
luminosity in the flow ($L/L_{\rm ph}\sim 20$). To find at what frequencies this
component is expected to show up, we can use $L_{\rm ph}$ to calculate the 
comoving temperature of the flow for a black-body spectrum through the
equation $L_{\rm ph}=16/3\sigma_{SB}r^2\Gamma^2T^4$. Solving for the 
temperature we have
\be
T(r_{\rm ph})=\frac{700}{ r_{\rm ph,11}^{7/12}}\frac{L_{52}^{1/4}}{(\epsilon
\Omega)_3^{1/12}{\sigma}_{0,2}^{1/2}} \qquad \rm eV.
\label{temp}
\ee

Setting again $ {\sigma}_{0}=100,\quad {\Omega}=10^4$ rad/sec,
$L=3.2\cdot10^{51}$ erg/sec/steradian, and $\xi=2\cdot10^{-3}$, we find
that the photospheric temperature in the central engine's frame is $T_{\rm ce}=
\Gamma T\simeq 60$ keV. So, we see that the photospheric emission peaks
at similar frequencies to  what is typically observed for the non-thermal emission. 

At this point it should be emphasized that for a comoving temperature 
of the order of a few hundred eV at the photosphere, the approximation of
a black-body spectrum emitted at that location may well be poor. 
This is because at that temperature electron scattering 
greatly dominates over free-free absorption and cannot in 
principle be neglected. The same concerns may be raised when
the photospheric emission is computed in the `standard' fireball
model.

\section{Conclusions}     

In this paper we have explored the spectra expected from Poynting-flux
dominated outflows and their relevance to GRBs. The field strength in such
outflows is much higher than in the usual internal shock models (much above
`equipartition' with the radiating fast particles).

We have used the `AC' model,
with geometrical configuration similar to the striped wind (Coroniti 1990) in which
the magnetic field varies with a typical wavelength of $\lambda\sim 2\pi
c/\Omega$. The dynamics of this outflow were studied in detail in Papers I, 
II where it was shown that almost half of the magnetic energy released through
magnetic dissipation accelerates the flow to large bulk Lorentz
factors. The other half of the magnetic energy accelerates the
particles of the flow, mainly above the Thomson photosphere and can
serve to power the prompt emission.

An important property of energy dissipation by magnetic reconnection is that the
residence time of the accelerated electrons in the reconnection layer is long
compared with the synchrotron energy loss time scale. The electrons are thus in
approximate balance between heating and cooling, and a sudden acceleration
assumption as usually made in (internal) shock acceleration is not valid. In a
magnetic model, the break in the Gamma-ray spectrum is therefore unlikely to be
associated with the cooling spectrum of suddenly accelerated particles. 

We have shown that a GRB-like Gamma-spectrum with a break can be produced by 
reconnection in a Poynting-flux dominated outflow if the dissipated magnetic 
energy heats a small fraction ($\sim 10^{-4}$) of the electron population. 
A better understanding of relativistic reconnection will be needed to determine
if this number is realistic.

In the magnetic model for GRB emission presented here X-ray 
flashes are interpreted to belong to the same family as GRBs and are the result 
of an outflow starting with a smaller magnetization parameter $\sigma$
(higher baryon loading) than needed to produce a GRB.
 X-ray flashes can also be interpreted within the fireball model with some
modification of the parameters.
              
Finally we have explored the consequences of the strong synchrotron absorption
due 
to the high magnetic fields strengths in a Poynting-flux dominated outflow. We find 
that it must be important around the electron scattering photosphere,  and will result 
in harder spectra in the soft X-ray range.         

\begin{acknowledgements}
Giannios acknowledges partial support from the EC Marie Curie Fellowship HPMTCT
2000-00132 and the Program ``Heraklitos'' of the Ministry of Education of Greece.
\end{acknowledgements}

\end{document}